# Sentiment Analysis on the News to Improve Mental Health

Saurav Kumar
School of Medicine
Stanford University
Saratoga, United States
sauravkumr2022@gmail.com

Rushil Jayant
Student Body
Homestead High School
Cupertino, United States
rushil.jayant@gmail.com

Nihaar Charagulla
Student Body
Bellarmine College Preparatory
San Jose, United States
nihaar.charagulla22@bcp.org

*Abstract*— **The popularization of the internet created a revitalized digital media. With monetization driven by clicks, journalists have reprioritized their content for the highly competitive atmosphere of online news. The resulting negativity bias is harmful and can lead to anxiety and mood disturbance. We utilized a pipeline of 4 sentiment analysis models trained on various datasets – using Sequential, LSTM, BERT, and SVM models. When combined, the application, a mobile app, solely displays uplifting and inspiring stories for users to read. Results have been successful – 1,300 users rate the app at 4.9 stars, and 85% report improved mental health by using it.**

*Keywords—sentiment analysis, media, negativity, mental health*

## I. Introduction

With the popularization of hand-held and portable technology over the past decade, access to information and the internet is growing every day. Due to this trend in technology, physical newspapers have started becoming obsolete [1].

This trend has created a competitive atmosphere among news organizations as they fight search engine optimization and write headlines designed to get clicks [2]. From prioritizing bringing diverse and unopinionated news to readers across the United States to fighting algorithms to get clicks, this shift in focus has clouded the purpose of the news. Information is more accessible than ever, but access to the right information becomes less and less accessible by the day.

Another unintended consequence of this competitive digital atmosphere is a negative bias [3]. Because news sites need to prioritize clicks on headlines to monetize, their strategies have shifted. This bias also has detrimental effects on mental health and mood, which will be discussed further in Section 2.

Google's BERT and OpenAI's GPT-3 have been two major players in natural language processing (NLP) lately. The applications of these models will be elaborated on in Section 3.

In this study, we applied 5 different models trained various datasets. Technologies used include GPT-3, BERT, Keras Sequential, LSTM, and SVM. The pipeline was applied to the news to analyze sentiment and filter out positive news. Once we completed the model, we developed a proprietary iOS Application to display the positive news and a few other uplifting features. Thus far, the app has been successful, with the vast majority of users reporting a tangible difference in mood and mental health, which we will explore in the rest of this paper.

The paper is as follows: Section 2 will focus on the psychological benefits of good news as well as the negativity bias currently in the media. Section 3 will explore the development and data analysis of the sentiment analysis model. Section 4 will discuss the design and development of the iOS App that the model was used for. Finally, section 5 will cover the results and conclude the paper.

## II. The Psychology of Bad News

### A. Data Analysis

First, in order to prove the extent to which the positivity platform suggested is necessary, we performed data analysis on headlines from the past few years.

We used a dataset of a million headlines from ABC, an Australian news organization. Although the organization is Australian, the negativity bias is world-wide [3], and the USA is the most mentioned country in the dataset, mentioned 9799 times in the headlines alone. For the purposes of this study, we first shuffled the dataset of around 1,226,258 news headlines and used 10,000 items from the data at random. The data used spanned from September, 2019 to December, 2020

In order to quickly and effectively analyze these articles, we utilized the SVM model discussed in section 3. Figure 1 demonstrates that the negative headlines vastly outnumbered the positive headlines by a factor of 49.19%. Although the model is only around 78% accurate, it is an effective baseline to establish the problem at hand.

Figure 2 is from Richard Vogg, a student and author from Germany. In his analysis of the million headlines, he used an older dataset from 2006 to 2016, with 1,076,225 values, and used a different sentiment analysis model [4]. He found that the mean positivity of articles every month was exclusively negative, which proves that the claimed negativity bias in the news is indeed a present phenomenon.

## B. Psychological Impacts of the Bias

Furthermore, multiple psychological studies prove that the effects of negative news on readers and watchers can be detrimental to mental health and mood.

For example, a case study by the University of California, Irvine, found that following the Boston Marathon Bombing, the media played a large role in "broadcasting acute stress" to the masses [5]. Figure 3 displays the substantial increase in acute anxiety and its correlation to media coverage of the bombing.

## C. Figures

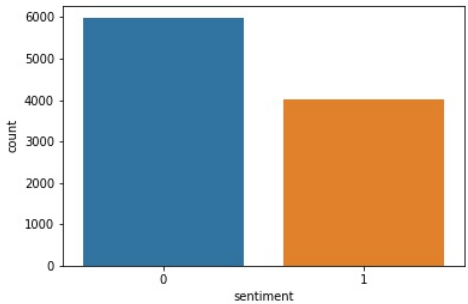

Fig. 1. Countplot of the sentiment of 10,000 headlines.

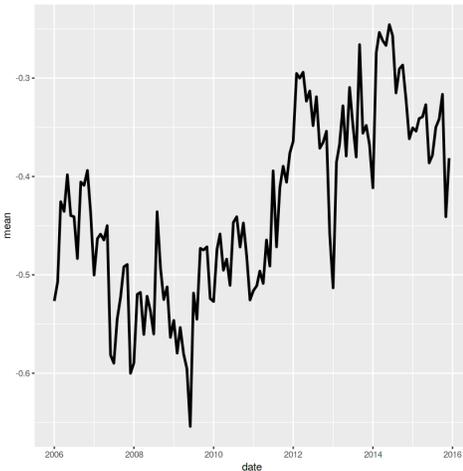

Fig. 2. Mean sentiment of headlines every month, 2006 to 2016 [4].

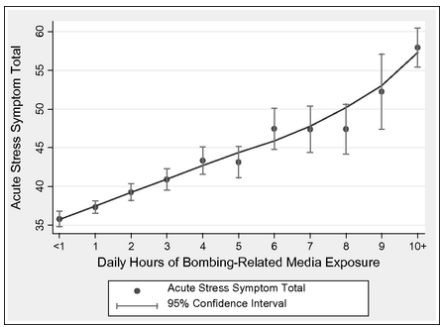

Fig. 3. Acute stress vs. Media coverage of bombing-related news [5].

## III. DEVELOPMENT OF THE PIPELINE

Because we are creating a pipeline for an app, and not for accuracy, we are optimizing for the effectiveness of filtering out only positive news. In other words, whenever there are errors in extracting sentiment, we want those errors to result in false negatives, rather than false positives. Therefore, we applied 5 models to this task in hopes of creating a layer of filters that no negative news could bypass.

## A. LSTM

In this model we used a Long Short Term Network model to train the dataset, and this fared much better than some of our other models. This model helps predict values in the dataset with increased accuracy. In fact, this model scored significantly better than other models at a 98% accuracy.

The dataset, which is also used in a few of the following models, was a compilation of 90,000 articles created by UCI [6]. Sentiment was flagged as a decimal value from -1 to 1, but we simplified that to include anything over 0 as 1 and anything below 0 as 0. This same modification was applied on all uses of this dataset.

Our LSTM model made use of multiple dense layers and dropout layers, increasing accuracy but more importantly preventing overfitting [7]. We had relu and sigmoid activation functions which added more diversity to the model.

We were impressed with the performance significantly and believe that it would be a viable tool for deployment. In our quest to use a multitude of models, this adds a unique diversity to our plethora of models we have. We believe that having more models, especially ones that use various different tools to predict values gives a more diverse insight into what goes on in real-life headlines.

## B. Sequential

Our sequential neural network was a very simple model utilizing the Keras API and the tensorflow tokenizer. With a 94% accuracy, we decided to use this model in our final pipeline.

A sequential model is comprised of multiple layers with single nodes, hence the name sequential [8]. Ours used 6 relu layers and 1 sigmoid layer. We also utilized the adam optimizer and the binary cross entropy loss function.

Ultimately, we were happy with the accuracy of this model as well as the fact that it was extremely light weight, processed inputs in milliseconds, and used the simple Keras API. Although not as interesting or novel as the other models, it was simple and easy to add to our pipeline.

## C. SVM

Our next model utilized sci-kit learn's Support Vector Machine (SVM) architecture. SVM is an algorithm that finds similarities between vectors and classifies the data. To do this, it finds places to create a hyperplane, or support vector, and inserts it [9].

To train this model, we opted to use a dataset of 1 million labelled tweets, instead of actual news headlines. This allowed for more diversity in interpretation and analysis. Since headlines are often very formal, they are restricted to certain words, which can lead to small chances of false negatives. However, twitter is extremely diverse, so we hoped a greater vocabulary would help the pipeline.

This model ended up being less accurate than the neural networks, at 77.9%, but it is lightweight and serves it purpose as a catch-through algorithm.

### D. BERT

BERT is a tool created by Google, as a bidirectional text reader trained on with over 110 million parameters [10]. BERT was heavily used in our GPT-3 algorithm to preprocess text. Essentially, we used the model to fine tune our analysis. It gives us the ability to preprocess the datasets and compare words with each other. BERT is especially helpful in tasks like Named Entity Recognition, Classification, and Question answering. For our purposes, we want our model to train on as much data there is available, and BERT achieves that goal.

The Amazon Product Review sentiment analysis dataset from Kaggle was used for the retraining of this model [11]. The input is vectorized text, and the output is a rating from 1 to 5, with 1 being negative and 5 being positive.

The model is very strict and ensures that there is no negative news passing through the filter. On our manually curated dataset of 200 news articles, it has a 100% accuracy on no negative news leaking through the filter, but discarded all but 20 articles including some positive ones. On a solely positive dataset, the model has a 72% accuracy, meaning it is very strict on all data, but that is acceptable and preferable for our conditions and use cases.

### E. GPT-3

GPT-3 is one of the most powerful algorithms available today. The power of GPT-3 comes behind all of its use cases, where developers can perform tasks like creating websites or writing entire blogs using GPT-3 just by asking it to do so with a single API call.

One June 10, 2021, OpenAI released GPT-3 "Fine-tuning" [12]. Fine-tuning entails tweaking a model to your custom needs. Previously, since GPT-3 is pretrained, customization was limited. To fine-tune a model for sentiment analysis, we created a customized dataset. We manually picked out and labeled 200 news headlines from Reddit. Through this, we went through 4 different models

#### a) Model 1 (using the UCI dataset).

Model 1 was trained on the UCI, news database collected with 90,000 values [6]. This database identified each news headline with a scoring metric between -1 and 1. Feeding 66,000 values into GPT-3, this model was largely overfitted, mainly marking headlines as negative. In our tests, Model 1 predicted inputs as negative 98% of the time, when only 60% of the training set was negative data.

#### a) Model 4 (using the hand-picked dataset).

Models 2 and 3 were variations of this, changing in news datasets, but the results were similar.

Since the majority of GPT-3's training revolved around text-based generation, it focuses on using and predicting text. Therefore reading 0's and 1's is less effective for GPT-3 than using text. In Model 4, we changed all outputs from "0's" to "negative" and "1's" to "positive". Even then, the model struggled to label positive values, predicting 91% of the testing data as negative.

In the end, using GPT-3 fine-tuning to detect positive news is less beneficial compared to alternative models, and thus we elected not to use this model. The key difference between fine-tuning GPT-3 and creating a BERT model comes with their original training data and operationality. BERT is a natural language processing model created with features like bidirectional reading for context, making it optimal for sentiment anlaysis. Conversely, GPT-3 is a generative model, so it is designed to simply predict the next word or phrase based on context, which is not optimal for sentiment analysis.

## IV. DEVELOPMENT OF THE PIPELINE

For the model to be implemented and interact with as many people as possible, we decided to utilize a mobile app. We named the platform Lapis News, after the gemstone Lapis Lazuli to represent positivity. Additionally, Lapis means "pencil" in Spanish, which represents the news aspect of the app.

For simplicity and efficiency, we chose to release the app solely on iPhones. To retain the best quality of the app, we developed the app natively with Swift.

Apple released SwiftUI, a new approach to frontend and UI programming in September of 2019 [13]. SwiftUI is optimized to be fast and easy to use with Swift, and it is fully programmatic instead of the drag-and-drop nature of UIKit. Due to the efficiency of this new library, we used SwiftUI to create the app's frontend.

### A. Feature Set

Since the ultimate goal of Lapis News was not simply to show users good news, but to make people smile and spread positivity, we decided on multiple features to include in the app. However, the flagship feature remained to be the positive news.

Firstly, we utilized animal videos. Animals can lighten moods, and many studies prove that pets and animals are a great way to improve mental health [14].

Secondly, we wanted to also insert motivational quotes. Quotes are a great way to spread knowledge and are efficiently chosen to have an impact on readers.

Finally, we chose to include jokes and puns to make users laugh.

### B. Design

Design is paramount in an app's design to capture users and deliver the best possible experience. It is even more crucial in positivity apps to complement the calming and uplifting nature of positivity [15].

Figure 4 depicts the current design of the app and all the relevant pages. The default color palette uses shades of light blue and gray, similar to the *lapis lazuli* stone.

Firstly, we can see the news page includes a list of headlines, formatted in cells with an image on the right. On a click, a modal view is presented to display the full article with details and citations.

Figure 4 also displays the animal videos scrolling feed. We decided on creating a spring-scrolling view similar to the popular video displaying pages of TikTok, Instagram Reels, and YouTube Shorts.

Figure 4's 4th mockup reveals the interface of the jokes and quotes page, and it is simply a cell with text, and then a horizontal spring-scrolling view to view different jokes. On a tap of the logo, the page becomes a quotes page and displays the quotes rather than the jokes.

Finally, the last mockup in Figure 4 is a brief look at the dark mode that the app features. It is similar to the default, but with a lighter shade of gray and a darker shade of blue.

### C. Development

Development was where we spent the most time on. Since we decided on using Swift, we had to optimize the software and ensure it worked well on all devices supported.

Firstly, the news feed was the most time-consuming task. Before we had our model working, we utilized a combination of semi-automatic curation and web scraping to display around 15 news articles every day. This feed was 100% accurate since it was curated manually and checked over to ensure that the news was uplifting and inspiring. However, once we completed the algorithm, we created a full pipeline to run once a day for news articles to feed through.

Figure 5 demonstrates our new algorithmic pipeline. First, we get headlines globally from our API. In order to cut out every negative article and only keep positive ones, we will run the articles through each of the 4 models we are using. The order is not very relevant to the pipeline, but our order is the following: Sequential, LSTM, SVM, BERT. However, since we are looking for uplifting news and not solely positive news, there will be some positive articles to cut out after the first two models finish their filter. Finally, we run the remaining articles through the BERT Model. The Lapis News app is currently published on the App Store for free download.

### D. Figures

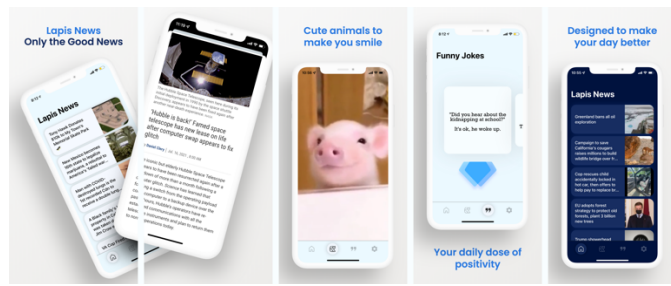

Fig. 4. Lapis News App Store Screenshots.

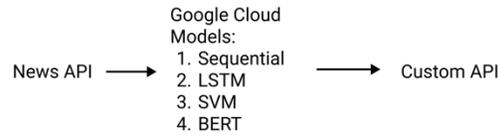

Fig. 5. Lapis News Backend Pipeline.

## V. CONCLUSION

As digital media grows, news will continue on its trend of negativity regardless of its effect on the mental health and wellbeing of readers. Information is important, and in an age with the internet enabling access to billions around the globe, it is unfortunate that the priority of news organizations has shifted from providing accurate and unbiased facts [2]. With our algorithmic pipeline, using technology like Sequential, LSTM, SVM, and BERT models, we created a sentiment analysis filter to separate positive articles from a daily bank of thousands of fresh headlines. Our app was carefully designed, features were thoughtfully selected, and the result was tuned and tweaked to our liking. The results say that the idea was a success, and there is market demand for the app. The app has 1,300 users and is growing rapidly. It has a 4.9 star rating on the App Store, and over 85% of users report a tangible increase in mood by using the app. However, our solution to this problem is not enough. For the negativity bias to disappear, news organizations must act and actively rethink their strategies and reprioritize their content.